\documentclass[12pt]{iopart}

\begin{document}

\title{Nonperturbative Green's function technique for nonequilibrium steady
state}

\author{Jongbae Hong}

\address{Department of Physics,
Seoul National University, Seoul 151-747, Korea \\
Max-Planck-Institut f\"ur Physik Komplexer Systeme, D-01187
Dresden, Germany} \ead{jbhong@snu.ac.kr}
\begin{abstract}

Nonperturbative dynamic theory has a particular advantage in
studying the transport in a quantum impurity system in a steady
state. Here, we develop a new approach for obtaining the retarded
Green's function expressed in resolvent form. We use the
Heisenberg picture to facilitate dynamic theory and propose a new
systematic method of collecting the basis vectors spanning the
Liouville space, which is the most crucial step in obtaining the
resolvent Green's function. We obtain all the linearly independent
basis vectors for studying the single-impurity Anderson models
with one and two reservoirs. The latter is an appropriate model
for studying the Kondo phenomenon in a steady state when a bias is
applied. This is one of long standing subjects in theoretical
condensed matter physics.

\end{abstract}



\maketitle

The single-particle Green's function is a basic tool for studying
the correlation effects in many-body systems. However, calculation
of this function for a strongly correlated system is not usually
successful because nonperturbative treatment is generally
required. Moreover, the situation becomes even worse for strongly
correlated systems under a steady state. Such systems are quantum
point contact~\cite{cronen,dicarlo,sfigakis,sarkozy}, single
electron transistor with a quantum
dot~\cite{cronen2,kogan,amasha,nygard}, and scanning tunneling
microscopy on a magnetic atom adsorbed on a metallic
surface~\cite{mad,mano,neel,otte}. These steady state Kondo
phenomena remain unexplored because no successful theory, which
can handle the nonequilibrium steady state for a quantum impurity
system with strong electron correlation, is available.

A well-known formula for a steady current through a small
interacting region connected to charge reservoirs having different
chemical potentials has been presented more than a decade
ago~\cite{meir,hersh}. The formula is given by
\begin{eqnarray} J=-\frac{e}{\hbar}
\int\frac{d\omega}{2\pi}\frac{\Gamma^L(\omega)
\Gamma^R(\omega)}{\Gamma^L(\omega)+\Gamma^R(\omega)}[f_L(\omega)-f_R(\omega)]\,
{\rm Im}G_{dd\sigma}^+(\omega) \label{current}
\end{eqnarray}
for proportional lead functions, i.e.,
$\Gamma^L(\omega)\propto\Gamma^R(\omega)$, where  $L$ and $R$
denote the left and right reservoirs, respectively and $f(\omega)$
denotes the Fermi distribution. The retarded Green's function
$G_{dd\sigma}^+(\omega)$ for the mediating atom must be obtained
for a given bias. However, this steady state Green's function has
never been obtained for an atom at which a strong interaction
exists.

The purpose of this study is to present a new method of
determining the steady state Green's function appearing in Eq.
(\ref{current}). For this purpose, we adopt the resolvent Green's
function in the Heisenberg picture~\cite{fulde}, which is given by
for a fermion system as
\begin{equation}
iG^{+}_{ij\sigma}(z)=\langle c_{i\sigma}|(z{\bf I}+i{\rm\bf
L})^{-1}|c_{j\sigma}\rangle, \label{heisenberg}
\end{equation}
where $z=-i\omega+\eta$ and $c_{i\sigma}$ indicates a fermion of
spin $\sigma$ annihilating at a state or site $i$. The symbols
${\rm\bf I}$ and ${\rm\bf L}$ in Eq. (\ref{heisenberg}) denote the
identity operator and the Liouville operator, respectively. The
latter is defined by ${\rm\bf L}{\hat{\cal O}}=[{\cal
H},{\hat{\cal O}}]$, where $\cal H$ and $\hat{\cal O}$ are the
Hamiltonian and an arbitrary operator, respectively. Although we
are familiar with the resolvent form in the Schr\"odinger picture,
\begin{equation}
G^{+}_{ij}(\omega)=\langle\phi_i|(\omega+i\eta-{\cal
H})^{-1}|\phi_j\rangle, \label{resolvent}
\end{equation}
on the basis of the static basis $\{\phi_j|j=0,1,\cdots,\infty\}$,
we employ Eq. (\ref{heisenberg}) using the operator basis in this
study because dynamical approach is more appropriate for handling
steady state situations.

Calculating the retarded Green's function in resolvent form starts
from ensuring a complete set of basis vectors. A notable point is
that the resolvent form does not provide information about the
basis vectors. Therefore, the Lanczos
algorithm~\cite{lanczos,dago} in the Sch\"odinger picture and the
projection operator technique~\cite{mori,zwan} in the Heisenberg
picture are used to obtain the static and dynamic basis vectors,
respectively. These techniques, however, require a heavy work for
a nontrivial system because they employ the Gram-Schmidt
orthogonalization procedure for the linearly independent vectors
${\cal H}^n|\phi_0\rangle$ or ${\bf L}^n c_{i\sigma}$, where $n=0,
1, \cdots, \infty$. For this reason, we propose a new methodology
that provides a simpler procedure to obtain the basis vectors.

The retarded Green's function $G_{dd\sigma}^+(\omega)$ can be
obtained by calculating the matrix inverse $({\rm\bf
M}^{-1})_{dd}$, where ${\rm \bf{M}}$ is composed of the elements
\begin{equation}
{\rm \bf{M}}_{ij}=z\delta_{ij}-\langle{i{\rm\bf L}\hat e}_j|{\hat
e}_i\rangle=z\delta_{ij}+\langle{\hat e}_j|i{\rm\bf L}{\hat
e}_i\rangle,  \label{matrixm}
\end{equation}
once a complete basis $\{{\hat e}_\ell|\ell=1,\cdots,\infty\}$
spanning the Liouville space is given. We choose ${\hat
e}_1=c_{d\sigma}$ to obtain $G_{dd\sigma}^+(\omega)$. The inner
product in Eq. (\ref{matrixm}) is defined by $\langle\hat e_k|\hat
e_\ell\rangle\equiv \langle\{\hat e_k,\hat
e_\ell^\dagger\}\rangle$, where the angular and curly brackets
denote statistical average and anticommutator, respectively. We
now present a systematic method to collect the basis vectors.

In the previous report~\cite{hong07}, we have obtained the basis
vectors intuitively. In this study, however, we provide a
systematic manner to secure universal validity and extensive
applicability. Expressing the resolvent Green's function in a
matrix form such as
\begin{equation}
iG_{ij\sigma}^+(z)= (\langle{\hat A}| \, \, \,\langle{\hat B}|)
\left(\begin{array}{cc} {\widehat G}_A \, \, \, \, \, \,  0\, \,
\\ 0 \, \, \, \, \, \, \, \,  {\widehat G}_B
\end{array} \right)\left(\begin{array}{c} |{\hat A}\rangle \\ |{\hat B}\rangle
\end{array} \right) \label{matrixform}
\end{equation}
is the most crucial step to obtain the basis vectors in systematic
manner. Once we get this form for the Green's function, it is
convinced that the Liouville space of $G_{ij\sigma}^+(z)$ will be
spanned completely by the linearly independent components of
vectors $|{\hat A}\rangle$ and $|{\hat B}\rangle$.

To obtain the matrix product form in Eq. (\ref{matrixform}), we
expand the Green's function operator ${\widehat G}=[(z{\bf
I}+i{\bf L}_I)+i{\bf L}_C]^{-1}$, where ${\bf L}_I$ and ${\bf
L}_C$ represent the Liouville operator for the isolated part of
the Hamiltonian ${\cal H}_I$ and the connecting part ${\cal H}_C$,
respectivel, in powers of ${\bf L}_C$ using the operator identity
$({\hat A}+{\hat B})^{-1}={\hat A}^{-1}-{\hat A}^{-1}{\hat
B}({\hat A}+{\hat B})^{-1}$. Then, one can express the retarded
Green's function as follows:
\begin{equation}
iG_{ij\sigma}^+(z)=\langle
c_{i\sigma}|\widehat{G}_{I}|c_{j\sigma}\rangle -\langle
c_{i\sigma}|\widehat{G}_{I}|i{\rm \bf
L}_{C}\widehat{G}_{I}c_{j\sigma}\rangle+\langle
c_{i\sigma}\widehat{G}_{I}i{\rm \bf L}_{C}|\widehat{G}|i{\rm \bf
L}_{C}\widehat{G}_{I}c_{j\sigma}\rangle, \label{expand2}
\end{equation}
where ${\widehat G}_I\equiv (z{\bf I}+i{\bf L}_I)^{-1}$. Equation
(\ref{expand2}) is transformed into a matrix form,
\begin{equation}
iG_{ij\sigma}^{+}(z)=\left( \langle c_{i\sigma}| \, \, \,
 \langle{\Phi_i}| \right) {\sf G} \left(
|c_{j\sigma}\rangle \, \, \, |{\Phi_j} \rangle \right)^T,
\label{green2}
\end{equation}
where ${\sf G}=\left(
\begin{array}{cc} {\widehat G}_I \, \, \,  -{\widehat G}_I\, \,
\\ 0 \, \,\, \, \, \, \, \, \, \, \, \,   {\widehat G}
\end{array} \right)$,
$|\Phi_{j} \rangle=|i{\bf L}_C{\widehat G}_{I}c_{j\sigma}\rangle$,
 and the superscript $T$ denotes the transpose.
 After a similarity transformation that diagonalizes ${\sf G}$,
 Eq. (\ref{green2}) is given by
\begin{equation}
iG_{ij\sigma}^{+}(z)=\left( \langle \widetilde{c}_{i\sigma}| \, \,
\,
 \langle{\Phi_i}| \right) {\sf G}_D \left(
|\widetilde{c}_{j\sigma}\rangle \, \, \, |\Phi_j\rangle \right)^T,
\label{green}
\end{equation}
where ${\sf G}_D=\left(
\begin{array}{cc} {\widehat G}_I \, \, \, \, \, \,  0\, \,
\\ 0 \, \, \, \, \, \, \, \,  {\widehat G}
\end{array} \right)$  and
\begin{equation}
|{\widetilde c}_{j\sigma}\rangle=|c_{j\sigma}\rangle+({\widehat
G}_{I}-{\widehat G})^{-1}{\widehat
G}_{I}|\Phi_j\rangle=|c_{j\sigma}\rangle+|{\rm\bf
L}_{C}^{-1}(-iz{\rm\bf I}+{\rm\bf L})\Phi_j\rangle
.\label{newvector}
\end{equation}
Equation (\ref{green}) is the form we want to have and the
linearly independent components of vectors $|{\widetilde
c}_{j\uparrow}\rangle$ and $|\Phi_{j}\rangle$ completely span the
Liouville space of $G_{ij\sigma}^{+}(\omega)$. In conclusion, the
systematic method for collecting the basis vectors involves
finding all linearly independent components comprising the vector
$|{\widetilde c}_{j\sigma}\rangle$ because $|\Phi_j\rangle$ is
contained in it.

We will subsequently demonstrate the determination of the basis
vectors with an example. We are interested in a mesoscopic system
with a mediating Kondo atom between two metallic reservoirs. This
system can be described by a single-impurity Anderson model with
two metallic reservoirs, whose Hamiltonian is written as
\begin{eqnarray}
{\cal
H}&=&\sum_{k,\sigma,\alpha=L,R}\epsilon_kc^{\alpha\dagger}_{k\sigma}
c^{\alpha}_{k\sigma}+ \sum_{\sigma}\epsilon_dc^\dagger_{d\sigma}
c_{d\sigma}+Un_{d\uparrow}n_{d\downarrow} \nonumber \\
&+&\sum_{k,\sigma,\alpha=L,R}(V_{kd}c^\dagger
_{d\sigma}c^\alpha_{k\sigma}+V^*_{kd}c^{\alpha\dagger}_{k\sigma}
c_{d\sigma}), \nonumber
\end{eqnarray}
where $\epsilon_{k}$, $\epsilon_{d}$, $V_{kd}$, and $U$ indicate
the energies of an electron of momentum $k$ in the lead, the level
of a mediating atom, the hybridization between the atom and lead,
and the on-site Coulomb repulsion at the atom, respectively.
Because the two-reservoir Anderson model is a straightforward
extension of the one-reservoir model, we first consider a
one-reservoir Anderson model. Its isolated and coupled parts are
given by
\begin{equation}
{\cal H}_I=\sum_{k,\sigma}\epsilon_kc^\dagger_{k\sigma}
c_{k\sigma}+\sum_\sigma\epsilon_dc^\dagger _{d\sigma}c_{d\sigma}+
Un_{d\uparrow}n_{d\downarrow}, \label{isol}
\end{equation}
and
\begin{equation}
{\cal H}_C = \sum_{k,\sigma}(V_{kd}c^\dagger
_{d\sigma}c_{k\sigma}+V^*_{kd}c^\dagger_{k\sigma}c_{d\sigma}).
\label{connect}
\end{equation}

We first demonstrate the calculation of $\Phi_d=i{\bf L}_C({\bf
I}+i{\bf L}_I)^{-1}c_{d\uparrow}$ and then $|{\widetilde
c}_{d\sigma}\rangle$ for ${\cal H}_I$ and ${\cal H}_C$ above. One
can easily see that $({\bf I}+i{\bf L}_I)^{-1}c_{d\uparrow}$
yields only two operators, $c_{d\uparrow}$ and
$n_{d\downarrow}c_{d\uparrow}$. Applying ${\rm\bf L}_{C}$ to these
operators changes the index $d (k)$ into $k (d)$ in fermion
operators and $n_{d\downarrow}$ into $j^-_{d\downarrow}$.
Therefore, the components of $\Phi_d$ are given by
\begin{equation}
\Phi_d=(c_{k\uparrow}, \hspace{0.3cm}
n_{d\downarrow}c_{k\uparrow}, \hspace{0.3cm}
j^-_{d\downarrow}c_{d\uparrow}), \label{phid}
\end{equation}
where $k=1, 2, \cdots, \infty$ and
$j^-_{d\downarrow}=i\sum_k(V_{kd}c^\dagger
_{d\sigma}c_{k\sigma}-V^*_{kd}c^\dagger_{k\sigma}c_{d\sigma})$. We
now focus on the vector $|{\widetilde c}_{d\sigma}\rangle$ in Eq.
(\ref {newvector}). The operator ${\bf
L}\Phi_d=[H,c_{k\uparrow}]+[H,n_{d\downarrow}c_{k\uparrow}]+[H,j^-_{d\downarrow}c_{d\uparrow}]$
yields $c_{d\uparrow}$ and $c_{k\uparrow}$ from the first
commutator, $n_{d\downarrow}c_{d\uparrow}$,
$n_{d\downarrow}c_{k\uparrow}$, and
$j^-_{d\downarrow}c_{k\uparrow}$ from the second commutator, and
$j^-_{d\downarrow}c_{d\uparrow}$,
$j^-_{d\downarrow}c_{k\uparrow}$,
$j^+_{d\downarrow}c_{d\uparrow}$,
$j^-_{d\downarrow}n_{d\downarrow}c_{d\uparrow}$, and
$j^+_{d\downarrow}n_{d\uparrow}c_{d\uparrow}$ from the third
commutator. The last operator
$j^+_{d\downarrow}n_{d\uparrow}c_{d\uparrow}$ is a vanishing one,
and the operator $j^-_{d\downarrow}n_{d\downarrow}c_{d\uparrow}$
is dynamically equivalent to $j^-_{d\downarrow}c_{d\uparrow}$.
Thus, the linearly independent components of $({\bf I}+{\bf
L})\Phi_d$ are classified into two groups, i.e., one involving
$c_{k\uparrow}$ and the other involving $c_{d\uparrow}$. They are
\begin{equation} ({\bf I}+{\bf L})\Phi_d^{\rm k}=(c_{k\uparrow}, \hspace{0.3cm}
n_{d\downarrow}c_{k\uparrow}, \hspace{0.3cm}
j^-_{d\downarrow}c_{k\uparrow}), \label{phiLd}
\end{equation}
for $k=1, 2, \cdots, \infty$, and
\begin{equation}
({\bf I}+{\bf L})\Phi_d^{\rm d}=(c_{d\uparrow}, \hspace{0.3cm}
n_{d\downarrow}c_{d\uparrow}, \hspace{0.3cm}
j^-_{d\downarrow}c_{d\uparrow}, \hspace{0.3cm}
j^+_{d\downarrow}c_{d\uparrow}). \label{phiLd2}
\end{equation}

As a last step, we apply ${\bf L}_C^{-1}$, which is equivalent to
the repeated application of ${\bf L}_C$, to the operators in Eqs.
(\ref{phiLd}) and (\ref{phiLd2}). One can easily see that the
multiple application of ${\bf L}_C$ to $c_{k\uparrow}$,
$c_{d\uparrow}$, and $n_{d\uparrow}$ simply reproduces the
existing operators and additionally
$j^+_{d\downarrow}c_{k\uparrow}$. Remaining linearly independent
operators are come from ${\bf L}_C^{n}j^\mp_{d\downarrow}$, where
$n=1, 2, \cdots, \infty$. Thus, one may classify the basis vectors
into two groups such as
\begin{equation}
{\bf L}_C^{-1}({\bf I}+{\bf L})\Phi_d^{\rm k}=\{c_{k\uparrow},
\hspace{0.3cm} n_{d\downarrow}c_{k\uparrow}, \hspace{0.3cm}
j^-_{d\downarrow}c_{k\uparrow}, \hspace{0.3cm}
j^+_{d\downarrow}c_{k\uparrow}, \hspace{0.3cm} ({\bf
L}_C^{n}j^\mp_{d\downarrow})c_{k\uparrow}\} \label{phiLd3}
\end{equation}
and
\begin{equation}
{\bf L}_C^{-1}({\bf I}+{\bf L})\Phi_d^{\rm d}=\{c_{d\uparrow},
\hspace{0.3cm} n_{d\downarrow}c_{d\uparrow}, \hspace{0.3cm}
j^-_{d\downarrow}c_{d\uparrow}, \hspace{0.3cm}
j^+_{d\downarrow}c_{d\uparrow}, \hspace{0.3cm} ({\bf
L}_C^{n}j^\mp_{d\downarrow})c_{d\uparrow}\}. \label{phiLd4}
\end{equation}
The operators shown in Eqs. (\ref{phiLd3}) and (\ref{phiLd4})
describe all possible linearly independent manners of annihilating
an up-spin at the site, $d$, of the mediating atom in time $t$.
They completely span the Liouville space of $c_{d\uparrow}(t)$ in
other words.

It is important to identify the meaning of the basis vectors
$({\bf L}_C^{n}j^\mp_{d\downarrow})c_{k\uparrow}$ and $({\bf
L}_C^{n}j^\mp_{d\downarrow})c_{d\uparrow}$ for later discussion.
Application of ${\bf L}_C$ to $j^\mp_{d\downarrow}$ gives rise to

$${\bf L}_Cj^-_{d\downarrow}=-i\sum_{\bf l}\sum_{\bf k}(V_{{\bf
l}d}V_{{\bf k}d}^* c^\dagger_{{\bf l}\downarrow}c_{{\bf
k}\downarrow}+V_{{\bf k}d}V_{{\bf l}d}^*c^\dagger_{{\bf
k}\downarrow}c_{{\bf l}\downarrow})+2i\sum_{\bf k}|V_{{\bf k}d}|^2
c^\dagger_{d\downarrow}c_{d\downarrow}$$ and
$${\bf L}_Cj^+_{d\downarrow}=\sum_{\bf l}\sum_{\bf k}V_{{\bf
l}d}V_{{\bf k}d}^* c^\dagger_{{\bf l}\downarrow}c_{{\bf
k}\downarrow}-\sum_{\bf l}\sum_{\bf k}V_{{\bf l}d}^*V_{{\bf k}d}
c^\dagger_{{\bf k}\downarrow}c_{{\bf l}\downarrow}.$$ These
operators represent a round trip of a down-spin electron between
the mediating atom and metallic lead. Therefore, $({\bf
L}_C^{n}j^\mp_{d\downarrow})$ depicts $n$-time trip of a down-spin
electron between the atom and lead without coming into contact
with an up-spin electron at the atom. This hardly happens in
reality.

Even though all linearly independent members of a basis of the
Liouville space for the single-reservoir Anderson model have been
obtained, working with these full members yields unnecessary
complication that cannot be easily overcome. One of the advantages
of the present method is that the physical meaning of each basis
vector is clear. Therefore, it is possible to eliminate some of
basis vectors that are unimportant in describing the dynamics of
the system at a particular parameter regime. A particular concern
must be given to the basis vector $n_{d\downarrow}c_{d\uparrow}$.
Since $n_{d\downarrow}^2=n_{d\downarrow}$, this basis vector
involves all higher orders of $U$-processes. Therefore, one must
eliminate $n_{d\downarrow}c_{d\uparrow}$ in studying Kondo regime
unless all $U$-processes are treated exactly. However,
$n_{d\downarrow}c_{d\uparrow}$ will be the most important basis
vector when we study a small $U$-regime. As mentioned above,
multiple round trips without coming into contact with other spin
electron at the mediating atom are practically rare; therefore,
one may further eliminate the basis vectors $({\bf
L}_C^{n}j^\mp_{d\downarrow})c_{d\uparrow}$. In other words, we
assume that the down-spin electron annihilates once it arrives at
the mediating atom. This assumption may be valid enough to
describe the spin exchange interaction in the Kondo regime.

Now, turn our attention to the basis vectors combined with $c_k$,
i.e., those shown in Eq. (\ref{phiLd3}). This part of the basis
vectors is transformed into the self-energy under matrix reduction
that is discussed later. We neglect the basis vectors $({\bf
L}_C^{n}j^\mp_{d\downarrow})c_{k\uparrow}$ by the same reason
neglecting $({\bf L}_C^{n}j^\mp_{d\downarrow})c_{d\uparrow}$
above. We also find that the role of the basis vectors
$j^\mp_{d\downarrow}c_{k\uparrow}$ is the same as that of
$n_{d\downarrow}c_{k\uparrow}$. However, their contribution to
constructing self-energy is negligible compared with that of
$n_{d\downarrow}c_{k\uparrow}$. Therefore, we further eliminate
$j^\mp_{d\downarrow}c_{k\uparrow}$. Then, we finally obtain the
reduce the Liouville space constructed by the basis vectors
$$(c_{d\uparrow}, \hspace{0.3cm}
j^-_{d\downarrow}c_{d\uparrow}, \hspace{0.3cm}
j^+_{d\downarrow}c_{d\uparrow}) \hspace{0.3cm} \mbox{\rm and}
\hspace{0.3cm} (c_{k\uparrow}, \hspace{0.3cm}
n_{d\downarrow}c_{k\uparrow}),$$ where $k=1, 2, \cdots, \infty$
for describing the dynamics of the single-impurity Anderson model
in the Kondo regime.

Since the resolvent form of $G_{dd\uparrow}^{+}(z)$ is given by
the Laplace transform of the coefficient $a_1(t)$ in the expansion
$c_{d\uparrow}(t)=\sum_{i=1}^\infty a_i(t){\hat e}_i$, where
${\hat e}_1=c_{d\uparrow}$, one must have an equality
$a_1(t)=G_{dd\uparrow}^{+}(t)$. This is valid when all the basis
vectors except $c_{d\uparrow}$ are orthogonal to $c_{d\uparrow}$.
This orthogonality condition is satisfied by introducing the
expression $\delta {\hat A}={\hat A}-\langle {\hat A}\rangle$ for
the operators $n_{d\uparrow}$ and $j^\mp_{d\uparrow}$. Thus, we
finally obtain the reduced Liouville space spanned by the basis
vectors
\begin{eqnarray*}(c_{d\uparrow}, \hspace{0.2cm} \delta
j^-_{d\downarrow}c_{d\uparrow}, \hspace{0.2cm} \delta
j^+_{d\downarrow}c_{d\uparrow}) \hspace{0.2cm} \mbox{\rm and}
\hspace{0.2cm} (c_{k\uparrow}, \hspace{0.2cm} \delta
n_{d\downarrow}c_{k\uparrow}), \label{basis}
\end{eqnarray*}
where $k=1, 2, \cdots, \infty$, for the one-reservoir Anderson
model in the Kondo regime. In practical calculations, we use the
normalized forms of the basis vectors. The basis vectors given
above are the same as those used in the previous
study~\cite{hong07} in which we obtained them intuitively.

The retarded Green's function $G_{dd\uparrow}^{+}(\omega)$ is
obtained by calculating the matrix inverse $({\rm\bf
M}^{-1})_{dd}$ constructed by the elements of Eq. (\ref{matrixm})
in terms of the basis given above. Calculating the inverse of an
infinite-dimensional matrix ${\rm\bf M}$ is a nontrivial problem.
For this purpose, we perform matrix reduction by using L\"owdin's
partitioning technique~\cite{loedin,mujika}. One can reduce the
infinite dimensional matrix ${\rm\bf M}$ to a $3\times 3$ matrix
for the basis given above. The reduced $\widetilde{{\bf
M}}^d_{3\times 3}$ is given by
\begin{eqnarray}
\widetilde{{\bf M}}^d_{3\times 3}&=&{\rm\bf M}^d_{3\times
3}-{\rm\bf M}_{\infty \times 3}({\rm\bf
M}_{\infty\times\infty}^{k})^{-1}{\rm\bf M}_{3\times\infty},
\label{eigen1}
\end{eqnarray}
where ${\rm\bf M}^d_{3\times 3}$ is the part constructed by the
basis vectors $ (c_{d\uparrow}, \, \, \delta
j^-_{d\downarrow}c_{d\uparrow}, \, \, \delta
j^+_{d\downarrow}c_{d\uparrow})$, ${\rm\bf
M}_{\infty\times\infty}^{k}$ by the basis vectors $(c_{k\uparrow},
\, \, \delta n_{d\downarrow}c_{k\uparrow}),  \, \,  \mbox{where}
\, \, \, \, k=1, 2, \cdots, \infty$, and ${\rm\bf M}_{\infty
\times 3}$ and ${\rm\bf M}_{3\times\infty}$ represent two corner
parts of the matrix ${\rm\bf M}$. The second term of Eq.
(\ref{eigen1}) constructs the self-energy of the retarded Green's
function given by $iG_{dd\uparrow}^{+}(\omega)=[(\widetilde{{\bf
M}}^d_{3\times 3})^{-1}]_{11}$. We will discuss the specific
results for the single impurity Anderson model in a separate
paper.

The Liouville space for the two-reservoir Anderson model is a
straightforward extension of that of the single-reservoir model.
The basis vectors are given by $$(c_{k\uparrow}^L, \, \, \delta
n_{d\downarrow}c_{k\uparrow}^L),  \, \, \mbox{where} \, \,  \, \,
k=1, 2, \cdots, \infty, $$ $$(\delta
j^{+L}_{d\downarrow}c_{d\uparrow}, \, \, \delta
j^{-L}_{d\downarrow}c_{d\uparrow}, \, \, c_{d\uparrow}, \, \,
\delta j^{-R}_{d\downarrow}c_{d\uparrow}, \, \, \delta
j^{+R}_{d\downarrow}c_{d\uparrow}),  \, \,  \, \, \mbox{and} $$
$$(\delta n_{d\downarrow}c_{k\uparrow}^R, \, \, c_{k\uparrow}^R),
\, \, \mbox{where} \, \,  \, \, k=1, 2, \cdots, \infty.$$ The
superscripts $L$ and $R$ denote the left and right metallic leads,
respectively. This basis is used in explaining the experiments for
the steady state Kondo phenomena occurring in quantum point
contact or quantum wire\cite{cronen,dicarlo,sfigakis,sarkozy},
single electron transistor\cite{cronen2,kogan,amasha,nygard}, and
scanning tunneling microscopy\cite{mad,mano,neel,otte}. The
reduced matrix in this case is a $5\times 5$ matrix, i.e.,
$\widetilde{{\bf M}}^d_{5\times 5}$, which is given by
\begin{equation}
\widetilde{{\bf M}}^d_{5\times 5}={\rm\bf M}^d_{5\times 5}-{\rm\bf
M}^L_{\infty \times 5}({\rm\bf
M}_{\infty\times\infty}^{kL})^{-1}{\rm\bf M}^L_{5\times\infty}
-{\rm\bf M}^R_{\infty \times 5}({\rm\bf
M}_{\infty\times\infty}^{kR})^{-1}{\rm\bf
M}^R_{5\times\infty},\label{eigen2}
\end{equation}
where ${\rm\bf M}^d_{5\times 5}$ is constructed by the basis
vectors $$ (\delta j^{+L}_{d\downarrow}c_{d\uparrow},
\hspace{0.3cm} \delta j^{-L}_{d\downarrow}c_{d\uparrow},
\hspace{0.3cm} c_{d\uparrow}, \hspace{0.3cm} \delta
j^{-R}_{d\downarrow}c_{d\uparrow}, \hspace{0.3cm} \delta
j^{+R}_{d\downarrow}c_{d\uparrow})  \, \, \, \, $$ and ${\rm\bf
M}_{\infty\times\infty}^{kL}$ denotes the $\infty\times\infty$
matrix constructed by the basis vectors $ (c_{k\uparrow}^L, \, \,
\delta n_{d\downarrow}c_{k\uparrow}^L), \, \, \mbox{where} \, \,
\, \, k=1, 2, \cdots, \infty$. The second and third terms of Eq.
(\ref{eigen2}) give rise to the self-energy of the retarded
Green's function, $iG_{dd\uparrow}^{+}(\omega)=[(\widetilde{{\bf
M}}^d_{5\times 5})^{-1}]_{33}$, of the two-reservoir Anderson
model.

This is the end of formalism for the retarded Green's function via
the new approach. However, calculating the matrix elements is
another laborious process to obtain the retarded Green's function
in a steady state. We present those calculations in a separate
work. Even though one calculate the matrix elements, a barrier for
the retarded Green's function still remains because the elements
of the matrix $\widetilde{{\bf M}}^d_{5\times 5}$ contain the
undetermined quantities $\langle n_{d\sigma}\rangle, \langle
j^{-L,R}_{d\sigma}\rangle$, and $\langle
j^{+L,R}_{d\sigma}\rangle$, which are determined by the retarded
Green's function. They are expressed by $G_{dd\sigma}^{+}(\omega)$
as follows:
\begin{eqnarray}
\langle n_{d\sigma}\rangle=-\frac{1}{\pi}\int_{-\infty}^\infty
\frac{f_L(\omega)\Gamma^L(\omega)+ f_R(\omega)\Gamma^R(\omega)}
{\Gamma^L(\omega)+\Gamma^R(\omega)}\,{\rm
Im}\,G_{dd\sigma}^{+}(\omega)d\omega, \, \, \label{nave}
\end{eqnarray}
\begin{eqnarray}
\langle j^{-L}_{d\sigma}\rangle
=-\frac{1}{\pi}\int\frac{d\omega}{2}[f_L(\omega)-f_R(\omega)]\widetilde{\Gamma}(\omega)\,{\rm
Im}\,G_{dd\sigma}^{+}(\omega) =-\langle j^{-R}_{d\sigma}\rangle,
\nonumber \\ && \label{current2}
\end{eqnarray}
and
\begin{eqnarray}
\langle j^{+L(R)}_{d\sigma}\rangle=
\int_\infty^\infty\frac{d\omega}{\pi}f_{L(R)}(\omega)\Gamma^{L(R)}(\omega)
{\rm Re}\,{G}_{dd\sigma}^+(\omega), \label{jplus}
\end{eqnarray}
where $\widetilde{\Gamma}(\omega)=\Gamma^L(\omega)
\Gamma^R(\omega)/[\Gamma^L(\omega)+\Gamma^R (\omega)]$. The
expressions of Eqs. (\ref{nave}) and (\ref{current2}) are valid
only when $\Gamma^L(\omega)\propto\Gamma^R(\omega)$. Because of
these formal relations, the retarded Green's function and the
above quantities must be obtained by running a self-consistent
loop \vspace{0.3cm}

 $\langle n_{d\downarrow}\rangle^{(0)}, \langle j^{\mp
L,R}_{d\downarrow}\rangle^{(0)} \rightarrow
{G}_{dd\uparrow}^{+(0)}(\omega) \rightarrow
\rho^{(0)}_\uparrow(\omega) \rightarrow \langle
n_{d\uparrow}\rangle^{(0)}, \langle j^{\mp
L,R}_{d\uparrow}\rangle^{(0)} \rightarrow \\ \indent
{G}_{dd\downarrow}^{+(0)}(\omega) \rightarrow
\rho^{(0)}_\downarrow(\omega) \rightarrow \langle
n_{d\downarrow}\rangle^{(1)}, \langle j^{\mp
L,R}_{d\downarrow}\rangle^{(1)} \rightarrow
{G}_{dd\uparrow}^{+(1)}(\omega) \rightarrow
\rho^{(1)}_\uparrow(\omega) \rightarrow \cdots$

\vspace{0.3cm} \noindent in terms of Eqs.
(\ref{nave})-(\ref{jplus}) and
$iG_{dd\sigma}^{+}(\omega)=[(\widetilde{{\bf M}}^d_{5\times
5})^{-1}]_{33}$.

In sum, we have presented a systematic methodology for selecting
basis vectors spanning the Liouville space, which is the most
crucial step in calculating the retarded Green's function in
resolvent form, and have suggested a procedure for calculating the
retarded Green's function. The method presented in this study has
several advantages: (i) one can find a complete set of basis
vectors systematically; (ii) the physical meanings of the basis
vectors obtained by this method are very apparent, allowing one to
identify and remove the unimportant basis vectors for a particular
parameter regime and construct a reduced Liouville space; and
(iii) the self-consistent calculation does not require a
nonequilibrium density matrix, which makes nonequilibrium problem
difficult. As a final comment, we mention on the remaining
unknowns in addition to $\langle n_{d\sigma}\rangle, \langle
j^{-L,R}_{d\sigma}\rangle$, and $\langle
j^{+L,R}_{d\sigma}\rangle$ in the matrix elements. These unknowns
may be used as free parameters characterizing the system under
consideration. In a separate study, we present the spectral
function and differential conductance for a specific system
showing steady state Kondo phenomenon.

\ack The author thanks J Yi and S H Yoon for valuable discussions.
This work was supported by the Korea Research Foundation Grant
funded by the Korean Government (KRF-2007-614-C00005).

\end{document}